\documentclass[12pt,aps,prl,onecolumn,showpacs,superscriptaddress,amsmath,amssymb]{revtex4-2}

\usepackage{setspace}

\usepackage{graphicx}
\usepackage{dcolumn}
\usepackage{bm}
\usepackage{amsmath}
\usepackage{amssymb}
\usepackage{graphicx}
\usepackage{indentfirst}
\usepackage{booktabs}
\usepackage{multirow}
\usepackage{colortbl}
\usepackage{epsfig}
\usepackage[colorlinks=true, linkcolor=red, citecolor=blue, urlcolor=red]{hyperref}

\begin{document}

\onehalfspacing

\title{Field-free perpendicular magnetization switching by altermagnet with collinear spin current}

\author{M. Q. Dong}
\affiliation{Key Laboratory of Computational Physical Sciences (Ministry of Education), Institute of Computational Physical Sciences, State Key Laboratory of Surface Physics, and Department of Physics, Fudan University, Shanghai 200433, China}

\author{Zhi-Xin Guo}
\email{zxguo08@xjtu.edu.cn}
\affiliation{State Key Laboratory for Mechanical Behavior of Materials, School of Materials Science and Engineering, Xi’an Jiaotong University, Xi’an, Shanxi 710049, China}

\author{Xin-Gao Gong}
\email{xggong@fudan.edu.cn}
\affiliation{Key Laboratory of Computational Physical Sciences (Ministry of Education), Institute of Computational Physical Sciences, State Key Laboratory of Surface Physics, and Department of Physics, Fudan University, Shanghai 200433, China}

\date{\today}

\begin{abstract}

The generation of collinear spin current (CSC), where both the propagation direction and spin-polarized direction aligned perpendicularly to the applied charge current, is crucial for efficiently manipulating systems with perpendicular magnetic anisotropy used in high-density magnetic recording. However, the efficient generation of CSC remains a challenge. In this work, based on the symmetry analysis, we propose that CSC can be effectively generated using altermagnets when the charge current is aligned along specific directions, due to spin-dependent symmetry breaking. This proposal is supported by density functional theory (DFT) and Boltzmann transport equation (BTE) calculations on a series of altermagnetic materials, including RuO$_2$, Mn$_5$Si$_3$, KRu$_4$O$_8$ and CuF$_2$, where unusually large CSC is produced by the charge current along certain orientations. Furthermore, we introduce a physical quantity, the spin-splitting angle, to quantify the efficiency of CSC generated by the charge current. We find that the spin-splitting angle ranges from 0.24 to 0.57 in these altermagnets, which is significantly larger than the spin-Hall angle typically observed in the anomalous spin-Hall effect, where the spin-Hall angle is generally less than 0.1. Our findings provide an effective method for manipulating spin currents, which is advantageous for the exploration of altermagnetic spintronic devices with field-free perpendicular magnetization switching.  

\end{abstract}

\maketitle


\section{\label{sec:introduction}introduction}

One of the most well-known applications of spintronics is electrically controllable nonvolatile magnetic random-access memory (MRAM) \cite{wolf2001spintronics, bhatti2017spintronics}. For early MRAM devices, electrical writing in multilayer structures such as spin valves and tunnel junctions is realized by injecting spin-polarized electrons from a "pinned" ferromagnet into a "free" ferromagnetic layer through a nonmagnetic insulating layer, inducing the so-called spin transfer torque (STT) \cite{slonczewski1996current, PhysRevB.54.9353, ralph2008spin, brataas2012current, khvalkovskiy2013basic}. However, since the spin-polarized electrons pass directly through the insulating layer, this writing process does not meet the requirements for low power consumption and high speed. The spin Hall effect (SHE), which arises from spin-orbit coupling (SOC), allows an in-plane charge current to generate an out-of-plane spin current with in-plane spin polarization, thereby exerting a torque on the adjacent ferromagnetic (FM) layer \cite{RevModPhys.91.035004, PhysRevB.95.014403, chernyshov2009evidence}. This effect, referred to as spin-orbit torque (SOT), overcomes the limitations of STT and has garnered significant attention over the past decade \cite{dyakonov1971current, PhysRevLett.83.1834, PhysRevLett.85.393, PhysRevLett.92.126603, murakami2003dissipationless, PhysRevLett.98.156601, mihai2010current, pi2010tilting, miron2011perpendicular, suzuki2011current}. Nevertheless, the SOT mechanisms for switching a perpendicularly magnetized system are non-deterministic \cite{PhysRevLett.109.096602, miron2011perpendicular}, which significantly limits their commercial applications, as materials with perpendicular magnetic anisotropy are essential for data storage media \cite{moser2002magnetic}.

To manipulate perpendicular magnetization deterministically, an external assisting magnetic field \cite{PhysRevLett.109.096602, miron2011perpendicular} or effective field \cite{yu2014switching, fukami2016magnetization, van2016field, lau2016spin, cai2017electric} along the current direction have been proposed. However, from the perspective of commercial viability, it is essential to develop new techniques for perpendicular magnetization switching that do not rely on external magnetic or effective fields (field-free switching). Recently, the anomalous spin-Hall effect (ASHE) has been discovered, which can generate spin currents with both the propagation direction and the spin-polarized direction aligned perpendicularly to the applied charge current (denoted as \textit{collinear spin current}, CSC), thereby enabling field-free switching \cite{baek2018spin, liu2020two, PhysRevB.100.184402, song2020coexistence, macneill2017control, PhysRevB.96.054450, shi2019all, kao2022deterministic, li2023intrinsic}. Nevertheless, the ASHE is generally much weaker than the SHE \cite{macneill2017control, PhysRevLett.106.036601, liu2012spin}, leading to a very small spin-Hall angle for the collinear spin current. This limitation necessitates a large current density to achieve deterministic switching, which does not align with the requirements for low power consumption. Moreover, the spin non-conserving nature of SOC, which allows for spin-current generation, simultaneously reduces the spin diffusion length, typically to the nanoscale, thereby further constraining the practical application of SOT \cite{RevModPhys.91.035004}. Consequently, there is a pressing need to explore new micromechanics that can effectively generate collinear spin currents for the advancement of high-performance MRAM devices.

Recently, a novel magnetic phase termed altermagnetism has been both theoretically proposed and experimentally verified \cite{PhysRevX.12.031042, PhysRevX.12.040501, PhysRevX.12.021016, PhysRevX.12.040002, feng2022anomalous, PhysRevB.107.L100418, PhysRevLett.132.176702, ma2021multifunctional, vsmejkal2020crystal, PhysRevLett.126.127701}. Unlike conventional collinear antiferromagnets, where opposite spin sub-lattices are connected through spatial translation ($\tau$) or spatial inversion symmetries ($\mathcal{P}$), altermagnets link the sub-lattices through crystal-rotation (proper or improper) symmetries ($\mathcal{R}$) \cite{PhysRevX.12.031042, PhysRevX.12.040501, PhysRevX.12.021016, PhysRevX.12.040002}. This unique symmetry results in spin-momentum locking (spin splitting) in momentum space, which generates a spin current when an external electric field is applied, similar to the SHE but without the involvement of SOC \cite{PhysRevLett.126.127701}. This characteristic introduces a new spin-splitting torque (SST) mechanism that combines the advantageous features of both STT and SOT, presenting significant potential for applications in low-power MRAM devices. However, to date, only SST based on non-collinear spin currents (where the spin current and spin polarization are perpendicular to each other) has been discovered, which still does not enable deterministic switching of perpendicular magnetization \cite{PhysRevLett.128.197202, fan2024robust}.

In this work, we propose that by manipulating the direction of the N\'{e}el vector in altermagnets, a collinear spin current can be effectively obtained without SOC, enabling high-efficiency deterministic switching of perpendicular magnetization. We first demonstrate the feasibility of producing such a CSC through a theoretical analysis of several altermagnets. Subsequently, using density functional theory (DFT) and Boltzmann transport equation (BTE) calculations, we reveal the unusually high efficiency of the CSC generated in altermagnets, which significantly surpasses that produced by the ASHE. This finding paves the way for the development of high-performance, low-power MRAM devices.

\section{\label{sec:result}result}
Firstly, we illustrate the micromechanics of generating spin current through spin splitting in altermagnets. In general, the symmetric operations of non-relativistic spin groups can be expressed as $[\mathbf{R}_i||\mathbf{R}_j]$ \cite{PhysRevX.12.031042,litvin1974spin, litvin1977spin}, where the element on the left side ($\mathbf{R}_i$) of the double vertical bar is a spin space operation, and the element on the right side ($\mathbf{R}_j$) denotes a real-space crystallographic operation. For collinear spin arrangements in crystals, the spin group includes the symmetric operation $[E||\mathcal{T}]$, where $E$ is the identity operation on the spin-space coordinates ($\mathbf{s}$) and $\mathcal{T}$ flips the sign of the crystal momentum ($\mathbf{k}$). Consequently, one can derive the relation: $[E||\mathcal{T}] \epsilon(\mathbf{k},\mathbf{s}) = \epsilon(-\mathbf{k},\mathbf{s})$, and thus,
\begin{eqnarray}
	\epsilon(\mathbf{k},\mathbf{s}) = \epsilon(-\mathbf{k},\mathbf{s})
	\label{eq:ek_coll_mag}
\end{eqnarray}
for all non-relativistic collinear magnets, since $[E||\mathcal{T}]$ is a symmetric operation applicable to non-relativistic collinear spin arrangements in crystals \cite{PhysRevX.12.031042}. For traditional collinear antiferromagnetic (AFM) materials, the opposite spin sub-lattices are connected through translation $[C_2||\tau]$ or inversion $[C_2||\mathcal{P}]$ symmetries (where $C_2$ transforms spin $\mathbf{s}$ into $-\mathbf{s}$). Since $\tau \epsilon(\mathbf{k},\mathbf{s}) = \epsilon(\mathbf{k},\mathbf{s})$ and $\mathcal{P} \epsilon(\mathbf{k},\mathbf{s}) = \epsilon(-\mathbf{k},\mathbf{s})$, it follows that $[C_2||\tau] \epsilon(\mathbf{k},\mathbf{s}) = \epsilon(\mathbf{k},-\mathbf{s})$ and $[C_2||\mathcal{P}] \epsilon(\mathbf{k},\mathbf{s}) = \epsilon(-\mathbf{k},-\mathbf{s})$ . Combining these with Eq. (\ref{eq:ek_coll_mag}), we can derive:
\begin{eqnarray}
	\epsilon(\mathbf{k},\mathbf{s}) = \epsilon(\mathbf{k},-\mathbf{s}),
	\label{eq:ek_coll_antimag}
\end{eqnarray}
which indicates that the energy bands of different spins are degenerate at every point in the Brillouin zone. As a result, when an external electric field is applied, the electrons of different spins respond identically to the electric field, which naturally does not produce spin-polarized currents.

In contrast, for altermagnets, the situation is different, as their sub-lattices are linked by crystal-rotation (proper or improper) symmetries, i.e.,  $[C_2||C_n]$ or  $[C_2||\textbf{\textit{m}}]$ ($C_n$ represents $n$-fold ($n > 1$) rotation operation, and $\textbf{\textit{m}}$ represents spatial mirror operation). Note that the spatial translation operation $\tau$ is disregarded here, as it has no effect on the band structure. Considering that $C_n \epsilon(\mathbf{k},\mathbf{s}) = \epsilon(C_n\mathbf{k},\mathbf{s})$ and $\textbf{\textit{m}} \epsilon(\mathbf{k},\mathbf{s}) = \epsilon(\textbf{\textit{m}} \mathbf{k},\mathbf{s})$, one can obtain $[C_2||C_n] \epsilon(\mathbf{k},\mathbf{s}) = \epsilon(C_n\mathbf{k},-\mathbf{s})$ and $[C_2||\textbf{\textit{m}}] \epsilon(\mathbf{k},\mathbf{s}) = \epsilon(\textbf{\textit{m}} \mathbf{k},-\mathbf{s})$, we can derive:
\begin{eqnarray}
	\epsilon(\mathbf{k},\mathbf{s}) = \epsilon(\mathbf{k}^\prime,-\mathbf{s}),
	\label{eq:ek_coll_altmag}
\end{eqnarray}
for altermagnets, where $\mathbf{k}^\prime$ is obtained from $\mathbf{k}$ by $C_n$ or $\textbf{\textit{m}}$ operations. This means that a state with spin $\mathbf{s}$ at general position $\mathbf{k}$ in the Brillouin zone must correspond to a state with spin $-\mathbf{s}$ at another position $\mathbf{k}^\prime$ in the Brillouin zone. Consequently, electrons with different spins can form distinct currents when an external electric field is applied in a specific direction, resulting in a spin-polarized current. It is important to note that the improper rotation in this context should not include the spatial inversion operation $\mathcal{P}$; otherwise, one cannot obtain the spin current induced by spin splitting, as is the case in conventional AFM materials \cite{PhysRevX.12.031042}.

\begin{figure}
	\centering
	\includegraphics[width=\linewidth]{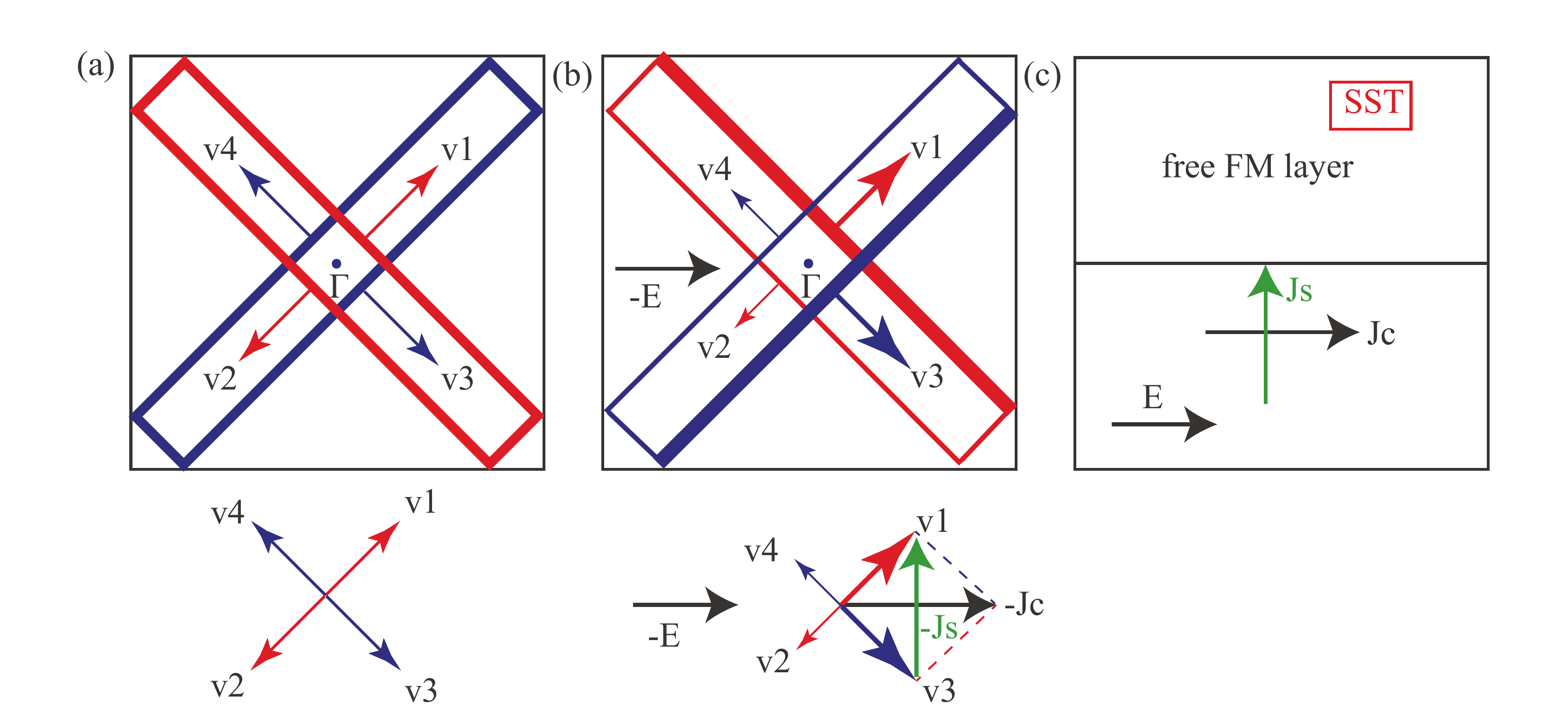}
	\caption{\label{fig:sst}(a) Simplified Fermi surface of $[C_2||C_{4z}]$ altermagnet in the absence of electric field. The red and blue lines represent the Fermi surfaces of spin-up and spin-down electrons, respectively. The red and blue arrows illustrate the Fermi velocity of spin-up electrons and spin-down electrons, correspondingly. (b) Redistribution of electrons at the Fermi surface induced by the application of an electric field, indicated by variations in line thickness. The green arrows depict the associated spin currents. (c) Schematic diagram illustrating the generation of collinear spin current (CSC, $J_s$) under external electric field and CSC applied spin-splitting  torque (SST) to the adjacent ferromagnetic layer.}
\end{figure}

In the absence of an external electric field, i.e., under equilibrium conditions, the electrons are symmetrically distributed in momentum space, and their velocities ($\textbf{\textit{v}}_{n\mathbf{k}} =\partial \epsilon_n(\mathbf{k})/\partial \mathbf{k}$) along all directions cancel each other out. As a result, the integral of the velocity of all electrons is zero. Consequently, the integral of the spin times the velocity of all electrons ($\int {\mathbf{s} \textbf{\textit{v}} \: d^3 \mathbf{k} }$) also equals zero, indicating that there is no macroscopic charge current or spin current, as illustrated in Fig. \ref{fig:sst}(a). On the other hand, when an external electric field is present, the distribution function of electrons near the Fermi level experiences a slight shift in the opposite direction of the electric field [see Fig. \ref{fig:sst}(b)]. At this point, the electron velocities along all directions can no longer cancel out. As a result, in addition to the longitudinal charge current, a net transverse spin current is also generated by the external electric field, as indicated in Fig. \ref{fig:sst}(b). In particular, in altermagnetic materials, when the N\'{e}el vector is perpendicular to the external electric field, the electric field (e.g., in the $x$-direction) will induce a transverse CSC (e.g., in the $y/z$ direction). This CSC can produce a significant SST, which is injected into the adjacent ferromagnetic (FM) layer to achieve high-efficiency deterministic switching of perpendicular magnetization, as shown in Fig. \ref{fig:sst}(c). 

We further discuss the electronic transport properties of altermagnets, which directly correlate to the characteristic of band splittings discussed above. Here, a spin conductance $\sigma_{\alpha \beta}^{\gamma\uparrow} (\sigma_{\alpha \beta}^{\gamma\downarrow})$ is defined, where $\alpha$ represents the propagation direction of spin current, $\beta$ means the direction of external electric field, and $\gamma$ is the direction of spin polarization, respectively. Note that “$\uparrow$” and “$\downarrow$” represent the spin-up and spin-down electrons, respectively. In altermagnets exhibiting symmetric operations such as $[E||C_{2z}]$ or $[E||C_{2z}\tau]$, one can derive the relation: $[E||C_{2z}] \epsilon(k_x,k_y,k_z,\mathbf{s}) = \epsilon(-k_x,-k_y,k_z,\mathbf{s})$ and thus $\epsilon(k_x,k_y,k_z,\mathbf{s}) = \epsilon(k_x,k_y,-k_z,\mathbf{s})$ by combining with Eq. (\ref{eq:ek_coll_mag}). In other words, the band structures of both spin-up and spin-down electrons exhibit $\textbf{\textit{m}}_z$ mirror symmetry with respect to the momentum space of the $k_z=0$ plane. Therefore, two of the three transverse components of electronic conductivity, namely, $\sigma_{zx}^{\gamma\uparrow} (\sigma_{zx}^{\gamma\downarrow})$ and $\sigma_{zy}^{\gamma\uparrow} (\sigma_{zy}^{\gamma\downarrow})$, must be zero, and only $\sigma_{yx}^{\gamma\uparrow} (\sigma_{yx}^{\gamma\downarrow})$ can be non-zero, where $\gamma=x,y,z$, representing the direction of spin polarization (see section SI and SII of Supplementary Material for details \cite{supplementary}).

\nocite{PhysRevB.59.1758, PhysRevB.50.17953, PhysRevB.54.11169, PhysRevB.47.558, PhysRevLett.77.3865}
\nocite{PhysRev.52.191, RevModPhys.84.1419, PhysRevB.56.12847, MOSTOFI2008685, Pizzi_2020}
\nocite{Ziman_1972, grosso2013solid, PIZZI2014422}

For convenience in discussing the physics of CSC, we denote the direction of the external electric field (or charge current) as the $x$-direction ($E_x$). Then, a spin current transporting along $\alpha$-direction ($\alpha=x,y,z$) with spin polarization along $\gamma$-direction ($\alpha=x,y,z$), i.e., $J_{\alpha x}^{\gamma}$, can be generally defined as \cite{PhysRevB.73.035323}
\begin{eqnarray}
	J_{\alpha x}^{\gamma}=(\sigma_{\alpha x}^{\gamma\uparrow}-\sigma_{\alpha x}^{\gamma\downarrow})E_x.
	\label{eq:js}
\end{eqnarray}
Note that when the external electric field is oriented along other directions (e.g., $E_y$ and $E_z$), one can also obtain non-zero CSC using a similar analytical approach. As shown in Eq. (\ref{eq:js}), a non-zero $(\sigma_{\alpha x}^{\gamma\uparrow}-\sigma_{\alpha x}^{\gamma\downarrow})$ implies that when an electric field is applied in the $x$-direction, a non-zero spin current in the $y$-direction can be generated, i.e., $J_{y x}^{\gamma} \ne 0$. It is noted that the spin polarization of most conduction electrons in a magnetic material is parallel to the magnetization direction, i.e.,  the N\'{e}el vector in altermagnets \cite{PhysRevLett.128.197202, fan2024robust}. Hence, a sizable non-zero CSC of $J_{y x}^{y} (J_{z x}^{z})$ can be achieved when the N\'{e}el vector is oriented along the $y$-direction ($z$-direction)\cite{PhysRevLett.128.197202}, provided by $\sigma_{y x}^{y\uparrow}-\sigma_{y x}^{y\downarrow}\ne 0 \ (\sigma_{z x}^{z\uparrow}-\sigma_{z x}^{z\downarrow}\ne 0)$.

Accordingly, for altermagnets exhibiting symmetric operations such as $[C_2||\textbf{\textit{m}}_z]$ or $[C_2||\textbf{\textit{m}}_z\tau]$, one can derive the relation $[C_2||\textbf{\textit{m}}_z] \epsilon(k_x,k_y,k_z,\mathbf{s}) = \epsilon(k_x,k_y,-k_z,\mathbf{-s})$. This means that the band structure of spin-up electrons and spin-down electrons are connected via $\textbf{\textit{m}}_z$ mirror symmetry with respect to momentum space of $k_z=0$ plane. Specifically, this leads to the relations $\sigma_{yx}^{y\uparrow} = \sigma_{yx}^{y\downarrow}$ and $\sigma_{zx}^{z\uparrow} = -\sigma_{zx}^{z\downarrow}$ (section SII of Supplementary Material \cite{supplementary}). As a result, $J_{y x}^{y}=0$ but $J_{z x}^{z}\ne 0$, indicating that a non-zero CSC with both spin propagation and spin polarization directions along the $z$-direction can also be obtained when the N\'{e}el vector is oriented along the $z$-direction.

\begin{figure}
	\centering
	\includegraphics[width=\linewidth]{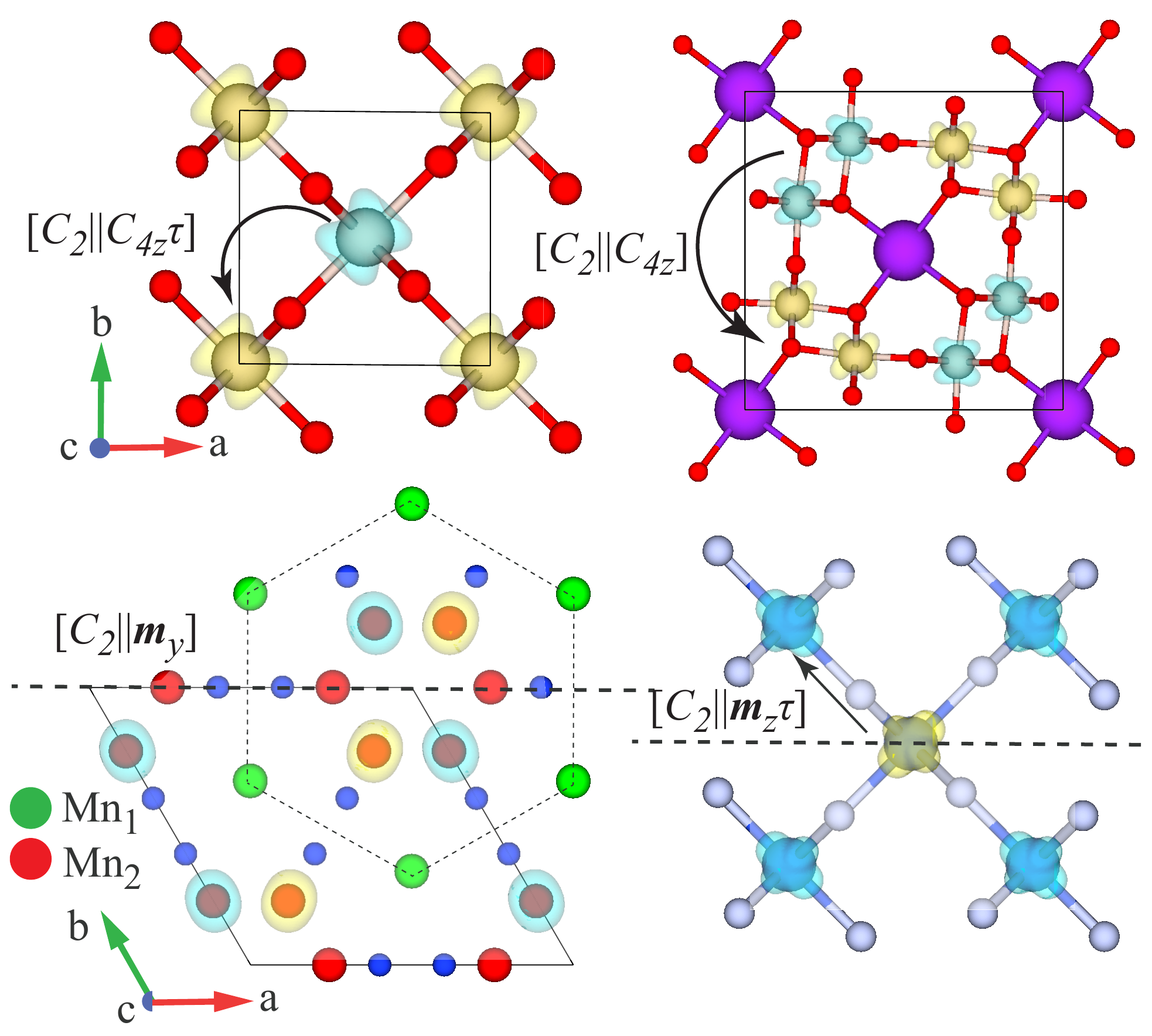}
	\caption{\label{fig:structure}Atomic structures, spin density distributions, and permitted altermagnetic symmetric operations for (a) RuO$_2$, (b) KRu$_4$O$_8$, (c) Mn$_5$Si$_3$ and (d) CuF$_2$.}
\end{figure}

To verify the theoretical derivations presented above, we further perform DFT calculations (see Supplementary Material \cite{supplementary} for details) on several metallic altermagnets, including RuO$_2$, KRu$_4$O$_8$ and Mn$_5$Si$_3$, which exhibit symmetric operation $[E||C_{2z}]$ or $[E||C_{2z}\tau]$. Additionally, we consider an insulating altermagnet, CuF$_2$ that has the symmetric operation $[C_2||\textbf{\textit{m}}_z\tau]$, for comparison.

The atomic structures of the four altermagnets are presented in Fig. \ref{fig:structure}. RuO$_2$ is a rutile oxide that belongs to the tetragonal $P4_2/mnm$ space group, consisting of two collinear opposite spin sub-lattices linked by a 4-fold crystal-rotation operation $[C_2||C_{4z}\tau]$ [Fig. \ref{fig:structure}(a)]. KRu$_4$O$_8$ features a hollandite structure that belongs to the centered tetragonal $I4/m$ space group and exhibits a collinear AFM ground state, with opposite sub-lattices connected by a 4-fold crystal-rotation operation $[C_2||C_{4z}]$ [Fig. \ref{fig:structure}(b)]. Mn$_5$Si$_3$ has a space group of $P6_3/mcm$, characterized by a hexagonal unit cell containing two formula units [Fig. \ref{fig:structure}(c)], i.e., four Mn atoms (Mn$_1$) located at the Wyckoff position 4d, and six Mn atoms (Mn$_2$) at the Wyckoff position 6g (two Mn$_2$ atoms are nonmagnetic) \cite{reichlova2024observation, gottschilch2012study, biniskos2023overview}. As for the CuF$_2$, it belongs to the monoclinic $P2_1/b$ space group consists of two collinear opposite spin sub-lattices linked by either a 2-fold crystal-rotation operation $[C_2||C_{2z}\tau]$ or mirror operation $[C_2||\textbf{\textit{m}}_z\tau]$ [Fig. \ref{fig:structure}(d)] \cite{PhysRevMaterials.8.034407,
billy1957crystal}. As shown below, although these altermagnetic materials have distinct atomic structures and belong different space groups, a common feature is that their band structures present significant band splitting along certain path in the Brillouin zone.

\begin{figure}
	\centering
	\includegraphics[width=\linewidth]{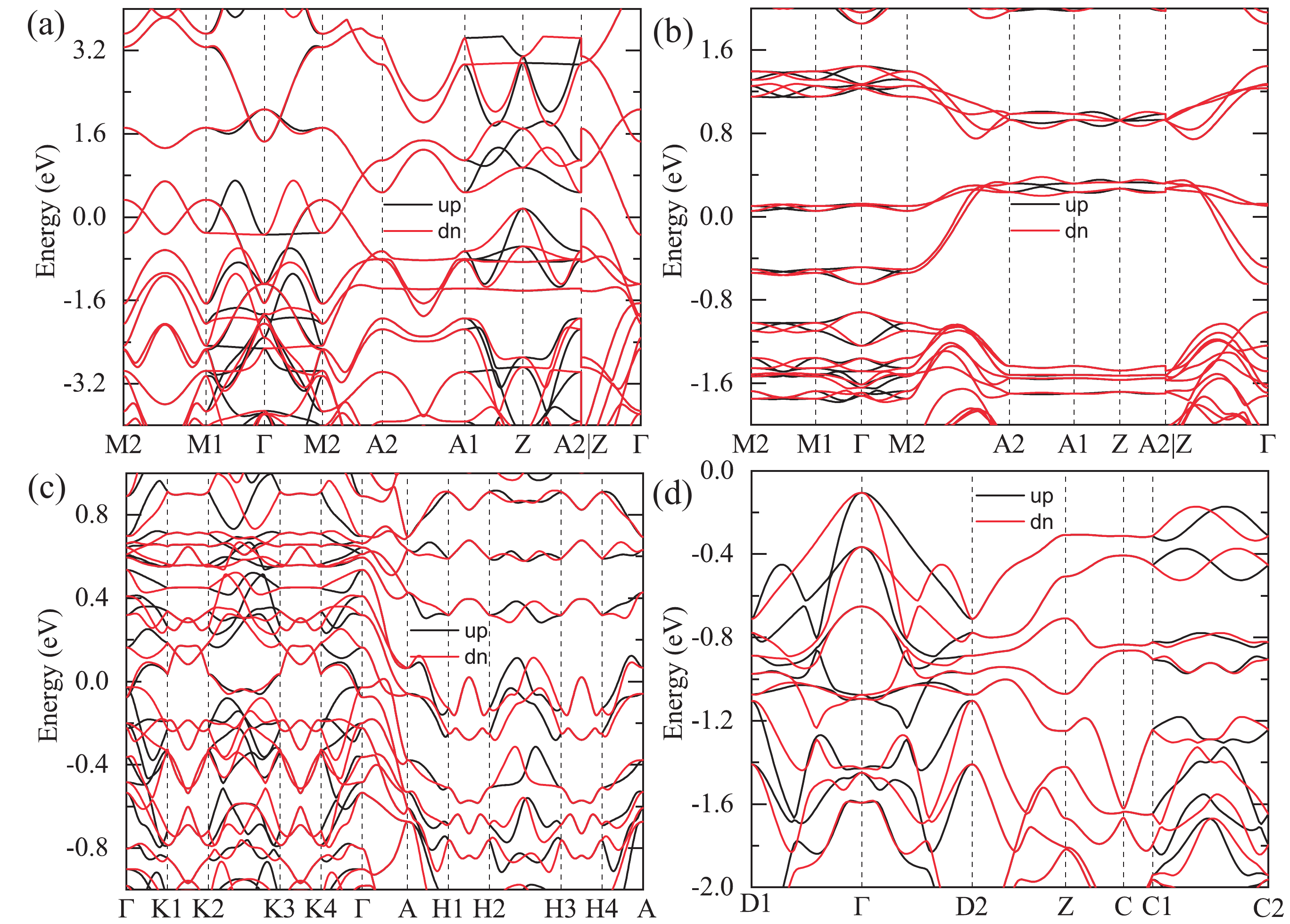}
	\caption{\label{fig:band}Band structures along high-symmetry paths for (a) RuO$_2$, (b) KRu$_4$O$_8$, (c) Mn$_5$Si$_3$ and (d) CuF$_2$. The shape of the Brillouin zone and the coordinates of the high-symmetry points are provided in the section SIII of Supplemental Materials \cite{supplementary}.
		}
\end{figure}

Figure \ref{fig:band} presents the DFT-calculated spin-polarized band structures of the four altermagnets discussed above. It is observed that the band structures of RuO$_2$ and KRu$_4$O$_8$ exhibit symmetric spin splitting along the paths M1-$\Gamma$-M2 and A1-Z-A2, while displaying spin degeneracy along the path Z-$\Gamma$ [Figs. \ref{fig:band}(a) and \ref{fig:band}(b)]. This result aligns well with the symmetry analysis based on the spin group operation $[C_2||C_{4z}\tau]$ or $[C_2||C_{4z}]$ for RuO$_2$ and KRu$_4$O$_8$. On the other hand, Mn$_5$Si$_3$ exhibits a symmetric operation $[C_2||\textbf{\textit{m}}_y]$. According to the symmetry analysis, its band structure should show spin splitting along the paths K4-$\Gamma$-K1 and K2-K3, while demonstrating spin degeneracy along the paths K1-K2 and K3-K4. As illustrated in Fig. \ref{fig:band}(c), the DFT-calculated band structure confirms these characteristics. As for CuF$_2$, which possesses a mirror operation of $[C_2||\textbf{\textit{m}}_z\tau]$, one can derive the relations: 
$[C_2||\textbf{\textit{m}}_z\tau] \epsilon(k_x,k_y,k_z\!=\!0,\mathbf{s}) = \epsilon(k_x,k_y,k_z\!=\!0,\mathbf{-s})$ and $[C_2||\textbf{\textit{m}}_z\tau] \epsilon(k_x,k_y,k_z\!=\!\pi,\mathbf{s}) = \epsilon(k_x,k_y,k_z\!=\!-\pi,\mathbf{-s}) = \epsilon(k_x,k_y,k_z\!=\!\pi,\mathbf{-s})$. This indicates that its band structure exhibits spin degeneracy in the Brillouin zone planes at $k_z=0$ and $k_z=\pi$. Therefore, we analyze the Brillouin zone plane at $k_z=\pi/2$ to illustrate its spin band-splitting characteristics, where band splitting is expected to occur along the D1-$\Gamma$-D2 and C1-C2 paths, respectively, in accordance with the mirror operation $[C_2||\textbf{\textit{m}}_z\tau]$ and the crystal-rotation operation $[C_2||C_{2z}\tau]$. As shown in Fig. \ref{fig:band}(d), this feature is also consistent with the DFT calculations.

\begin{figure}
	\centering
	\includegraphics[width=\linewidth]{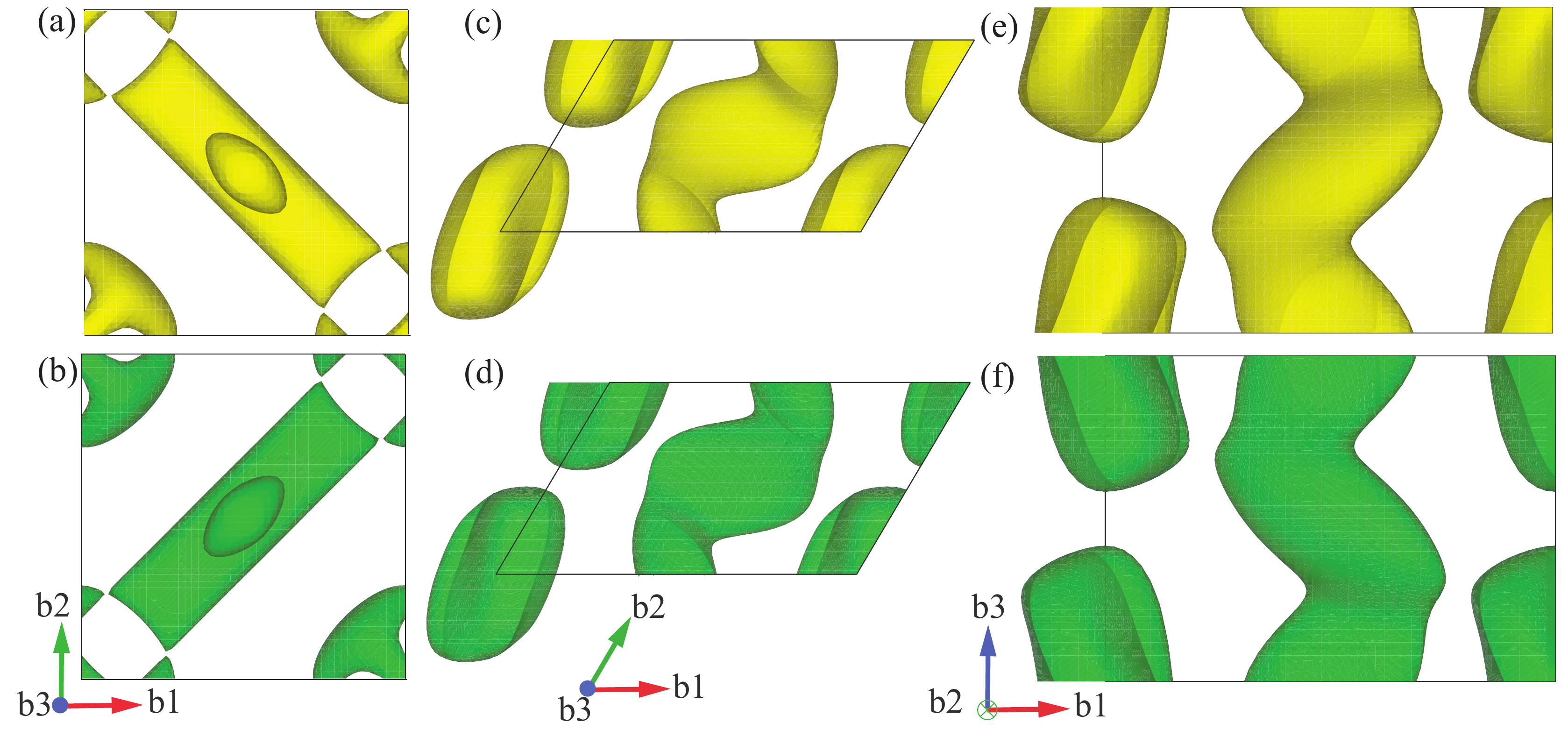}
	\caption{\label{fig:isosur}Isoenergetic surface of RuO$_2$ ($E=E_F$) and CuF$_2$ ($E=-0.27$ eV). (a) Top view of spin-up electrons of RuO$_2$. (b) Top view of spin-down electrons of RuO$_2$. (c) Top view of spin-up electrons of CuF$_2$. (d) Top view of spin-down electrons of CuF$_2$. (e) Side view of spin-up electrons of CuF$_2$ and (e) side view of spin-down electrons of CuF$_2$. Isoenergetic surface of KRu$_4$O$_8$ and Mn$_5$Si$_3$ are presented in the Supplemental Materials \cite{supplementary}.	
	}
\end{figure}

To more clearly illustrate the symmetry-correlated electronic structures of the altermagnets, we also calculate the isoenergetic surfaces of the four altermagnetic materials discussed above. As shown in Fig. \ref{fig:isosur}, the isoenergetic surfaces of RuO$_2$ (at $E=E_F=0$) and CuF$_2$ (at $E=-0.27$ eV) exhibit symmetry along certain directions in the Brillouin zone, which aligns well with the symmetry analysis discussed earlier. Note that in CuF$_2$ the non-zero energy level of isoenergetic surface is adopted, because it is an insulator [Fig. \ref{fig:band}(d)]. It is noteworthy that good agreement between the isoenergetic surface profiles and the symmetry analysis is also obtained in KRu$_4$O$_8$ and Mn$_5$Si$_3$ at $E_F$ (section SIV of Supplemental Materials \cite{supplementary}). Since the spin-splitting energy is generally at eV level, which is significantly greater than that induced by the SOC effect (typically at the meV level), the spin splitting in altermagnets is expected to generate a much more substantial CSC than that produced by ASHE. This makes altermagnets highly promising for applications in MRAM devices.

\begin{figure}
	\centering
	\includegraphics[width=0.8\linewidth]{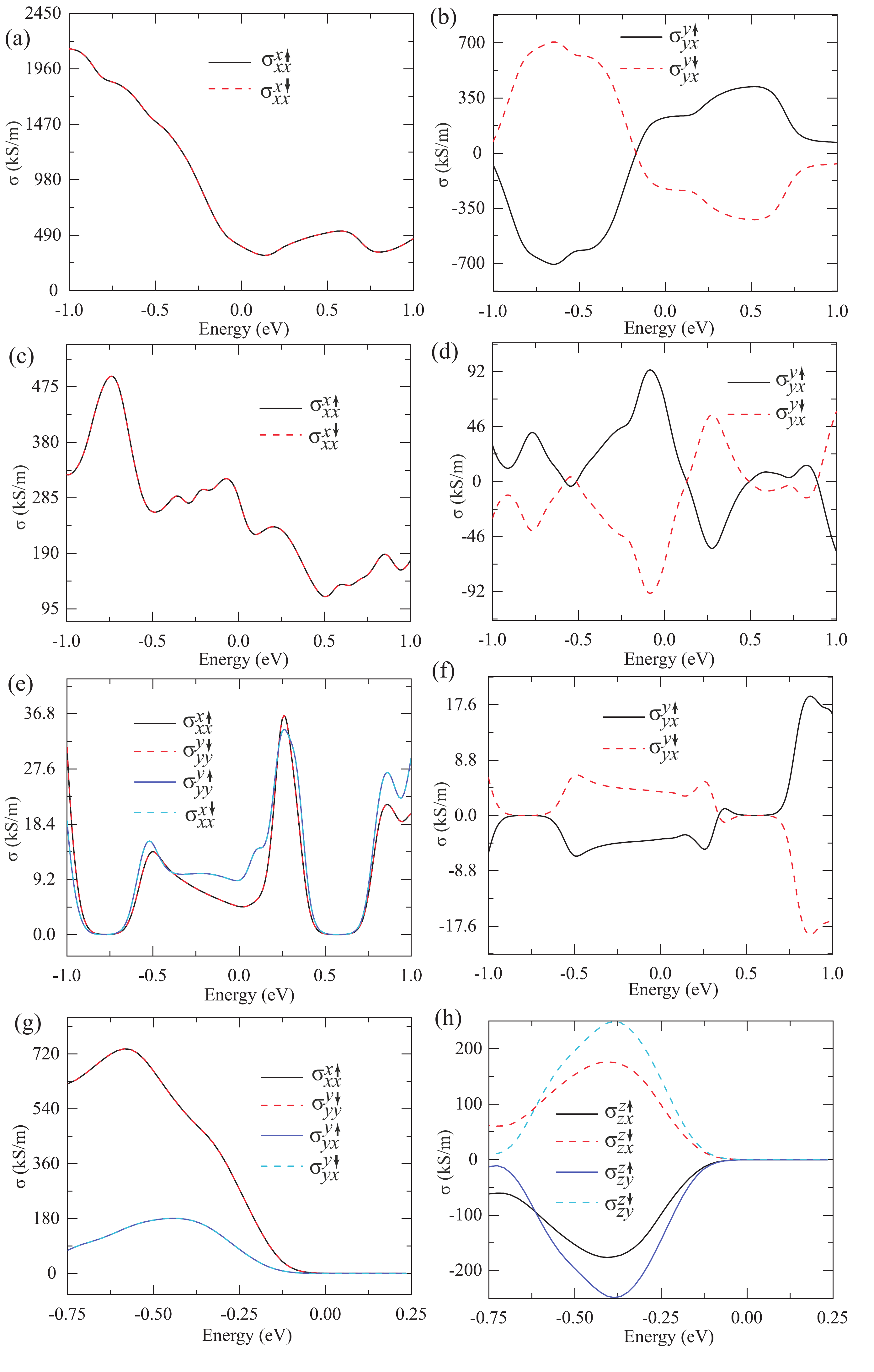}
	\caption{\label{fig:cond}Calculated conductivity for RuO$_2$ (a-b), Mn$_5$Si$_3$ (c-d), KRu$_4$O$_8$ (e-f), CuF$_2$ (g-h). For RuO$_2$, Mn$_5$Si$_3$ and KRu$_4$O$_8$, the calculated $\sigma_{zx}^{\uparrow/\downarrow}$ and $\sigma_{zy}^{\uparrow/\downarrow}$ are both zero for spin-up electrons and spin-down electrons.
	}
\end{figure}

We now turn to the discussion of the spin transport properties of the above four altermagnets. Figures \ref{fig:cond}(a) and \ref{fig:cond}(b) display the calculated longitudinal electric conductivity and transverse conductivity of RuO$_2$, with the charge current oriented along the [100] direction ($x$-direction) (see section SI of Supplementary Material for calculation details \cite{supplementary}). It is observed that spin-up electrons and spin-down electrons exhibit the same longitudinal conductivity but have opposite transverse conductivity. Consequently, a net longitudinal charge current $J_{xx}=(\sigma_{xx}^{\gamma\uparrow}+\sigma_{xx}^{\gamma\downarrow})E_x$ is generated in the $x$-direction, and a net transverse spin current $J_{yx}^{\gamma}=(\sigma_{xx}^{\gamma\uparrow}-\sigma_{zxx}^{\gamma\downarrow})E_x$ is produced in the [010] direction ($y$-direction). Since the spin polarization direction $\gamma$ is aligned along the N\'{e}el vector, a net CSC ($J_{yx}^{y}$) would obtained in RuO$_2$ when the N\'{e}el vector is along [010] direction.

In a manner analogous to the definition of the spin Hall angle \cite{RevModPhys.91.035004}, we further introduce a physical quantity, \textit{spin-splitting angle} ($\alpha$), to quantify the conversion efficiency from charge current to CSC resulting from spin splitting \cite{PhysRevLett.126.127701}. It is important to note that the spin-splitting angle is a tensor. For instance, for a CSC in the $y$-direction (the N\'{e}el vector is along $y$-direction) induced by electric field in $x$-direction, it can be expressed as:
\begin{eqnarray}
	\alpha_{yx}=\left|\frac{J_{yx}^{\gamma}}{J_{xx}}\right|
	=\left|\frac{\sigma_{yx}^{y\uparrow}-\sigma_{yx}^{y\downarrow}}
	{\sigma_{xx}^{y\uparrow}+\sigma_{xx}^{y\downarrow}}\right|.
	\label{eq:alpha}
\end{eqnarray}

From Figs. \ref{fig:cond}(a) and \ref{fig:cond}(b), one can observe a large spin splitting angle $\alpha_{yx}$ of approximately 0.57 at the $E_F$ for RuO$_2$. This indicates that the charge-to-spin transfer efficiency for the CSC is several dozen times greater than that produced via ASHE (0.013 \cite{macneill2017control}), highlighting the significant potential of altermagnets in SST-MRAM devices. It is noteworthy that the calculated transverse spin conductivity in the $z$-direction is zero for both spin-up electrons ($\sigma_{zx}^{z\uparrow}=0$) and spin-down electrons ($\sigma_{zx}^{z\downarrow}=0$), resulting in $\alpha_{zx}=0$. This result is also consistent with the symmetry analysis, i.e., the band structures of both spin-up and spin-down electrons exhibit $\textbf{\textit{m}}_z$ mirror symmetry. Additionally, according to the symmetry analysis, Mn$_5$Si$_3$ and KRu$_4$O$_8$ would exhibit a non-zero CSC and thus $\alpha_{yx}$ in the $y$-direction ([01$\bar{1}$0] for Mn$_5$Si$_3$ and [010] for KRu$_4$O$_8$) when the external electric field is applied along the [2$\bar{1}\bar{1}$0] and [100] directions (denoted as $x$-direction) for Mn$_5$Si$_3$ and KRu$_4$O$_8$, respectively. Figs. \ref{fig:cond}(c)-\ref{fig:cond}(f) further show the calculated longitudinal and transverse spin conductivities, demonstrating that  Mn$_5$Si$_3$ ($\alpha_{yx}=0.24$) and Mn$_5$Si$_3$ ($\alpha_{yx}=0.55$) also possess sizable spin splitting-angles.

\begin{table}[b]
	\caption{\label{tab:alpha}
		{Non-zero CSC ($J_s$) and spin-splitting angle ($\alpha$) in RuO$_2$, Mn$_5$Si$_3$, KRu$_4$O$_8$, and CuF$_2$. M and I denote metal and insulator, respectively. For RuO$_2$, Mn$_5$Si$_3$, KRu$_4$O$_8$, which are metals, we calculate $\alpha$ at the Fermi level (E=0).}
	}
	\begin{ruledtabular}
		\begin{tabular}{l*{2}{c}}
			Materials & $J_s$ & $\alpha$ \\
			\colrule 
			RuO$_2$ (M) 
			& 
			$\begin{pmatrix} 0 & J_{xy}^{x} & 0 \\ J_{yx}^{y} & 0 & 0 \\ 0 & 0 & 0 \end{pmatrix}$ 
			& 
			$\begin{pmatrix} 0 & 57\% & 0 \\ 57\% & 0 & 0 \\ 0 & 0 & 0 \end{pmatrix}$    \\
			Mn$_5$Si$_3$ (M)  
			& 
			$\begin{pmatrix} 0 & J_{xy}^{x} & 0 \\ J_{yx}^{y} & 0 & 0 \\ 0 & 0 & 0 \end{pmatrix}$ 
			& 
			$\begin{pmatrix} 0 & 24\% & 0 \\ 24\% & 0 & 0 \\ 0 & 0 & 0 \end{pmatrix}$    \\
			KRu$_4$O$_8$ (M) 
			& 
			$\begin{pmatrix} J_{xx}^{x} & J_{xy}^{x} & 0 \\ J_{yx}^{y} & J_{yy}^{y} & 0 \\ 0 & 0 & 0 \end{pmatrix}$ 
			& 
			$\begin{pmatrix} 32\% & 55\% & 0 \\ 55\% & 32\% & 0 \\ 0 & 0 & 0 \end{pmatrix}$    \\
			CuF$_2$ (I) 
			& 
			$\begin{pmatrix} 0 & 0 & J_{xz}^{x} \\ 0 & 0 & J_{yz}^{y} \\ J_{zx}^{z} & J_{zy}^{z} & 0 \end{pmatrix}$ 
			& insulator   \\		
		\end{tabular}
	\end{ruledtabular}
\end{table}

Finally, we present all the possible non-zero CSCs generated in the four altermagnets with the external electric filed varying among $x$ ([100]), $y$ (010), and $z$ ([001]) directions, as shown in Table \ref{tab:alpha}. For RuO$_2$, Mn$_5$Si$_3$ and KRu$_4$O$_8$, the CSCs in both the $x$ and $y$ directions can be effectively generated by appropriately choosing the direction of the charge current. In contrast, for CuF$_2$, CSCs can be obtained in any of the $x$, $y$, or $z$ direction. Combing with the unusually large spin splitting angles of these CSCs shown in Table \ref{tab:alpha}, these results present a promising way of realizing efficient deterministic switching using altermagnets. It is worth noting that, in addition to the CSCs propagating in the transverse direction of the charge current, a spin-polarized current in the longitudinal direction of the charge current can also be generated in KRu4O8, highlighting the rich spin-transport properties of altermagnets.

Before concluding, we would like to remark on the effect of SOC on the CSC. In the Supplemental materials, we present the calculated conductivity of RuO$_2$, both with and without the SOC effect. It is found that although the conductivity is slightly reduced in the presence of SOC, the positive and negative relationship between spin-up and spin-down electrons remains largely unchanged. As a result, the spin splitting angle $\alpha_{yx}$ remains substantial at 0.39, demonstrating the robustness of CSC in altermagnets.

\section{\label{sec:Conclusion}Conclusion}
In summary, based on the symmetry analysis, we find that the spin-dependent symmetry breaking can induce non-zero CSCs when the electric field is applied along specific directions in altermagnets, such as RuO$_2$, Mn$_5$Si$_3$, KRu$_4$O$_8$, and CuF$_2$. In use of DFT and BTE calculations, we further reveal that the CSCs in either transverse or longitudinal directions can be significantly large. We find that the efficiency of charge to collinear spin current (spin spin-splitting angle) ranges from 0.24 to 0.57 in these altermagnets, significantly larger than that achieved by the anomalous spin-Hall effect. Our findings provide a technique for effectively manipulating spin currents, which is advantageous for the exploration of altermagnetic spintronic devices with field-free perpendicular magnetization switching.


\section{\label{sec:acknowlodgment}acknowlodgments}
 
This work was supported by the Ministry of Science and Technology of the People’s Republic of China (2024YFA1408501), Natural Science Foundation of China (Grants No. 12474237), Science Fund for Distinguished Young Scholars of Shaanxi Province (Grant No. 2024JC-JCQN-09).

\bibliography{CSC}

\end{document}